\documentclass[aps,prl,onecolumn,superscriptaddress,groupedaddress]{revtex4}
\usepackage{subfig}
\usepackage{graphicx}  
\usepackage{dcolumn}   
\usepackage{bm}        
\usepackage{amssymb}   
\usepackage{slashed}
\usepackage{graphicx}				
\usepackage{amsmath}
\usepackage{mathtools}
\usepackage{tikz,pgf}
\usepackage{comment}
\usepackage{asymptote}
\usetikzlibrary{arrows,backgrounds}
\usetikzlibrary{fit,scopes,calc,matrix,positioning,decorations.pathmorphing}
\usepackage[all]{xy}
\usepackage{yfonts}
\newcommand{\bra}[1]{\ensuremath{\left\langle#1\right|}}
\newcommand{\ket}[1]{\ensuremath{\left|#1\right\rangle}}
\newcommand{\Bracket}[1]{\ensuremath{\left\langle#1\right\rangle}}

\begin{document}
\title{D-branes, AdS/CFT, dynamical Uhlmann Gauge, and stabilisation of a closed causal loop geometry}
\author{Andrei T. Patrascu}
\address{FAST Foundation, Destin FL, 32541, USA\\
email: andrei.patrascu.11@alumni.ucl.ac.uk}
\begin{abstract}
I show here that if we construct D-branes not in the form of infinite superpositions of string modes, in order to satisfy the technical condition of coherence by means of eigenstates of annihilation operators, but instead insist on an approximate but much more physical and practical definition based on phase coherence, we obtain finite (and hence realistic) superpositions of string modes that would form realistic D-branes that would encode (at least as a semiclassical approximation) various quantum properties. Re-deriving the AdS/CFT duality by starting in the pre-Maldacena limit from such realistic D-branes would lead to quantum properties on the AdS side of the duality. 
Causal structures can be modified in various many particle systems, including strings, D-branes, photons, or spins, however, there is a distinction between the emergence of an effective causal structure in the inner degrees of freedom of a material, in the form of a correlation generated effective metric for example in a spin liquid system, and the emergence of a causal structure in an open propagating system by using classical light. I will show how an Uhlmann gauge construction would add stability to a modified causal structure that would retain the shape of a closed causal loop.
Various other ideas related to the quantum origin of the string length are also discussed and an analogy of the emergence of string length from quantum correlations with the emergence of wavelength of an electromagnetic wave from coherence conditions of photon modes is presented. 
\end{abstract}
\maketitle
\section{Introduction}
The AdS/CFT correspondence [1] emerges naturally from string theory as a specific realisation of the general open-closed string duality [2]. This duality refers to the equivalence between two seemingly different descriptions of string interactions: the open string channel, describing strings with endpoints attached to D-branes that interact forming loop diagrams of open strings attached to boundaries, and the closed-string channel, in which the same physical process can equivalently be described by closed strings propagating freely without endpoints, exchanged between D-branes, and forming cylindrical diagrams. Formally this duality is expressed as an equality between two conformal field theory amplitudes 
\begin{equation}
Z_{open}(\tau)=Z_{closed}(\tilde{\tau})
\end{equation}
Here $\tau$ and $\tilde{\tau}$ are modular parameters related by a conformal transformation, essentially exchanging time and space coordinates in the string worldsheet. This relationship is fundamental and mathematically rigorous, arising naturally from modular invariance of the string worldsheet partition function. The validity of open-closed duality is rooted in the structure of two-dimensional conformal field theory and the geometry of string theory. String theory amplitudes are integrals over the moduli space of Riemann surfaces. Modular transformations map one amplitude with boundary (open strings) to amplitudes without boundary (closed strings). This ensures the mathematical equivalence between open and closed string descriptions. From a worldsheet perspective, consider a cylindrical worldsheet. In the open string channel the cylinder has boundaries (representing D-branes) and in the closed string channel the cylinder has no boundaries, as it is viewed as propagation of closed strings. Modular transformation maps one interpretation (boundaries) to the other (no boundaries), preserving the physics. In short, the reason open-closed duality holds is due to the underlying symmetries (modular invariance) and consistency conditions (unitarity and locality) of the string theory itself [3]. 
The AdS/CFT correspondence emerges naturally from the physics of D-branes in string theory [4]. Consider a stack of N coincident D3-branes embedded in Type IIB superstring theory. In the open-string description the dynamics of open strings with endpoints attached to the D3 branes generates a low energy theory of the brane worldvolume. At low energies, open strings attached to N D3 branes yield a gauge theory 
\begin{equation}
4D\;\; \mathcal{N}=4\;Super\;Yang-Mills\;(SYM)\; SU(N)
\end{equation}
In a closed string description, the strings propagate in the bulk, influenced by the mass-energy of the D-branes. At large $N$ and strong coupling, the geometry created by these branes can be described by supergravity, a low effective theory of closed strings. The near horizon geometry around the D3 branes is precisely 
\begin{equation}
AdS_{5}\times S^{5}
\end{equation}
Taking the low-energy limit (decoupling limit) isolates the physics on the D-branes from the asymptotically flat region. On the open string side (boundary theory) decoupling the gravitational interactions, the open string description becomes purely the boundary gauge theory
\begin{equation}
4D\;\mathcal{N}=4\;SU(N)\;SYM\;theory\;on\;flat\;spacetime
\end{equation}
On the closed string side (the bulk theory) near the horizon of the D3-branes, the closed string description simplifies to the supergravity low energy limit of closed strings on 
\begin{equation}
AdS_{5}\times S^{5}
\end{equation}
Thus we have two entirely equivalent descriptions of the same physical system: a gauge theory living on the boundary (open string origin) and a gravitational theory living in the bulk (closed-string origin). The equivalence is precisely the AdS/CFT correspondence 
\begin{equation}
Type\;IIB\;string\;theory\;on\;AdS_{r}\times S^{5} \Leftrightarrow 4D\;\mathcal{N}=4\;SU(N)\;SYM\;theory
\end{equation}
AdS/CFT is essentially a concrete manifestation of open-closed string duality [5]. Open string diagrams on D-branes can equivalently be described as closed-string diagrams exchanged between branes [6]. In the special limit where D-branes are numerous ($N\ll1$) their gravitational backreaction produces a well defined closed string geometry ($AdS_{5}\times S^{5}$). From a holographic standpoint, the open-string degrees of freedom on the brane correspond to closed-string gravitational dynamics in the bulk geometry. The branes become the boundary on which the CFT lives, while closed strings are the bulk degrees of freedom. This clearly shows why $AdS/CFT$ is obtained precisely from open-closed duality: it is nothing more than the realisation of this duality at large N and strong coupling [7]. 
The black brane interior however is not covered by the AdS/CFT duality [8]. Initially, the AdS/CFT correspondence has been stated as a duality between the bulk gravitational theory (closed string picture), usually classical supergravity or quantum gravity on an asymptotically AdS spacetime, and a boundary conformal field theory (open string picture), a non-gravitational quantum field theory defined on the boundary of AdS. The boundary theory lives at the spatial boundary (at infinity) of AdS spacetime. Thus, strictly speaking the duality directly encodes information accessible at or near the boundary and any event or phenomenon deep inside a black hole, appears at first to be beyond direct boundary reach. The naive holographic dictionary maps bulk fields near the boundary onto boundary operators, making it unclear how it directly describes the interior of the black hole. However, even though the AdS/CFT duality does not give an obvious dictionary mapping boundary operators directly to fields inside the black hole interior, it provides indirect methods to study the interior. Some of these are analytic continuation of boundary observables, quantum entanglement and entanglement wedges, reconstruction techniques a la HKLL reconstruction [9] or bulk operator reconstruction [10], state dependent operators [11] and complexity considerations, etc. In general, the black hole's interior is also encoded because the AdS/CFT is a non-local quantum holographic duality [12]. What seems like a spatially inaccessible region due to classical causal structure, becomes accessible when we notice that the duality maps entanglement and quantum information in a fundamentally non-local way. The boundary observables contain subtle quantum interference patterns, correlations, and entanglement structures that encode events behind the horizon. Although classical causality prevents still direct access, quantum correlations, like entanglement or complexity based reconstructions, provide indirect access to interior physics. In AdS/CFT the gauge theory lives very near the horizon of a black brane and is formed by the N D-branes linked by open strings. The bulk is the physics close to the horizon, described by classical gravity, hence closed strings, forming the AdS space, on the outside of the horizon. The bulk is then the region outside the horizon, farther away from the boundary where the gauge theory lives. 
Initially, when we derive the AdS/CFT correspondence starting from string theory and D-branes, gauge theory is a boundary theory originating from open strings attached directly to the stack of N D3-branes. In the low-energy limit, these open strings form a non-gravitational gauge theory ($\mathcal{N}$ SYM) that lives on the world-volume of the branes. When taking the low energy decoupling limit, the gauge theory effectively resides at the boundary at infinity of the emergent gravitational geometry, namely at the conformal boundary of the emergent AdS spacetime. For the closed strings (in the bulk theory) originally, the closed strings propagated everywhere, including near the branes. However, when we have a large number of branes ($N\rightarrow\infty$) and consider strong coupling, the branes' gravitational backreaction creates a non-trivial curved spacetime. In particular, close to the horizon, but still outside it, we will get the geometry we expected 
\begin{equation}
AdS_{5}\times S^{5}
\end{equation}
Thus the bulk in the standard AdS/CFT dictionary refers usually to this gravitational region near (but outside) the black brane horizon, extending all the way outwards to the boundary at infinity. This is the bulk from the viewpoint of AdS/CFT as usually presented. If however various methods for probing the inside of the black hole are employed, we can generalise and extend the meaning of the bulk. We can consider a two-sided eternal black AdS hole. In that case, the black hole is dual to two entangled copies of the CFT (a thermofield double state). In this case we would have an extended Penrose diagram. Here, the black hole interior is accessible via analytic continuation and entanglement wedge reconstruction, even though it lies technically behind event horizons. By definition, in this context the bulk is "extended" to mean also the interior, since it is part of the complete spacetime that emerges from quantum entanglement between boundary CFTs. Even in one-sided black holes formed by collapse, it is possible to reconstruct operators inside the horizon by using non-local, state dependent boundary operators. Therefore the notion of bulk is extended to mean that interior as well. Let us now describe Maldacena's limit in this context. Initially, in string theory setups we start with $N$ D3-branes placed in flat 10-dimensional spacetime. On the open strings side, the endpoints live directly on these D-branes, defining a gauge theory (e.g. $\mathcal{N}=4\;SYM$) on the brane world-volume itself. At this initial stage, the gauge theory clearly lives exactly where the D-branes are located. There is also no notion of a boundary at infinity yet. When we have many more D-branes ($N$ becomes large) that are stacked together, they strongly deform the spacetime geometry due to their collective gravitational attraction. The metric produced by a stack of $N$ D3-branes is given by 
\begin{equation}
\begin{array}{cc}
ds^{2}=\frac{1}{\sqrt{H(r)}}(-dt^{2}+d\vec{x}_{3}^{2})+\sqrt{H(r)}(dr^{2}+r^{2}d\Omega_{5}^{2}), & H(r)=1+\frac{L^{4}}{r^{4}}
\end{array}
\end{equation}
This geometry has a characteristic scale set by 
\begin{equation}
L^{4}\sim g_{s}N(\alpha')^{2}
\end{equation}
When we take the low energy decoupling limit (Maldacena limit) defined by sending $\alpha'\rightarrow 0$ (string length scale to zero) while keeping the energies fixed from the viewpoint of open strings on the branes, two crucial things happen. First, the open strings (the gauge theory side) stay confined exactly on the branes as they effectively decouple from gravity at infinity and form a purely field theoretic theory living precisely where the branes were placed originally (in the original coordinates, this is just flat space). Second, the closed strings (gravity side) will experience an infinitely strong "zoom-in" close to the horizon of the original D3-brane geometry. In other words we are looking closer and closer to the region right near the brane's horizon. Because we are zooming infinitely close to the brane, the horizon region expands by a very large amount. This near horizon region takes the form $AdS_{5}\times S^{5}$. This zooming-in near the horizon transforms the geometry from a brane localised in flat spacetime to an extended AdS space which now has a clear boundary at infinity. Before taking the limit, the branes are localised objects in flat space. After taking the limit, the near horizon region gets stretched infinitely creating a whole new geometry with an explicit boundary at infinity. The position of the gauge theory is coordinate dependent. In the original coordinates (flat embedding), gauge theory was localised exactly on the brane. In the near horizon coordinates, the region originally containing the brane (the original horizon of the brane) now appears infinitely far away, it becomes precisely the boundary at infinity of AdS. In other words, the brane location when viewed from inside the near horizon geometry is no longer close by. It moves to the conformal boundary of AdS. This is because the decoupling limit zooms infinitely into the region near the brane, stretching what used to be a finite location into an infinitely distant boundary. Finally, another aspect of the common language used in holography needs to be somehow clarified. When we hear in literature that we go "deep inside the bulk", sometimes this is referred to as the "quantum gravity limit", however, the AdS boundary is the high energy (UV regime of the gauge theory), while deeper inside the bulk we find the low-energy physics (IR regime of the gauge theory). In this context "deep inside the bulk" means going towards the interior of the AdS spacetime away from the boundary at infinity. One should notice that the AdS geometry itself arises strictly in the near-horizon limit, which eliminates the original asymptotically flat region (Minkowski spacetime) where the D-branes originally lived. In fact, the true full geometry originally was not just AdS, it was an interpolating geometry, going from flat Minkowski spacetime far from branes to AdS very close to the brains. Therefore quantum gravity, in a more fundamental sense (stringy effects, finite N corrections, gravitational quantum fluctuations, etc.) actually become relevant precisely in the regime where the Maldacena limit is no longer exact, i.e. precisely in the transitional region (near the original brane horizon), not strictly inside the AdS bulk geometry itself. The quantum gravitational phenomena are not really "deep inside AdS" but rather occur exactly in the domain where the original Maldacena limit is incomplete or starts to fail. Deep inside the bulk in the usual AdS/CFT terminology actually means going towards the classical IR gravitational regime, well described by classical (or semiclassical) gravity. Far from being "more quantum", this region is often less quantum, precisely because it's described by classical supergravity. True quantum gravitational effects lie precisely outside this simplified classical description in the region connecting classical AdS to flat Minkowski space, or near the UV boundary itself, where strongly quantum effects (such as stringy or brane excitations, finite N corrections, Planckian effects) become inevitable. 
With this clarified, let us have a look at the relationship between background space (target space), the D-branes moving within it, and the worldsheet dynamics of strings. The target space is the spacetime manifold in which strings propagate. Initially in the original D-brane construction we used to derive AdS/CFT, this target space is usually taken to be a simple flat Minkowski spacetime (10 dimensional Minkowski, e.g. type IIB superstring theory). D-branes themselves are extended objects occupying submanifolds of the target space. A D3-brane occupies a (3+1) dimensional submanifold of the original 10-dimensional spacetime. A completely separate 2-dimensional space (the sheet swept out by the string as it evolves in time) whose dynamics is described by a 2-dimensional conformal field theory is embedded in the target space and has its own dynamics. Strings propagating in target space correspond to fields defined on this worldsheet. This is the intrinsic space of string motion, and fields on the worldsheet encode positions of the string in target space. 
Let us start with a flat Minkowski space. Adding D-branes will change this. Many D-branes together carry energy and charge and thus by Einstein's equations (or supergravity equations) of motion, they curve spacetime around them. D-branes do not merely move in a fixed flat spacetime. They produce and respond to gravitational fields curving spacetime. Explicitly, for N D3-branes in type IIB string theory, the solution to supergravity equations gives a geometry
\begin{equation}
\begin{array}{ccc}
ds^{2}=H(r)^{-1/2}(-dt^{2}+d\vec{x}^{2}_{3})+H(r)^{1/2}(dr^{2}+r^{2}d\Omega_{5}^{2}), & H(r)=1+\frac{L^{4}}{r^{4}}, & L^{4}\sim g_{s}N(\alpha')^{2}
\end{array}
\end{equation}
This is precisely the geometry describing the target-space curvature created by the stack of D3 branes. The Maldacena limit ($\alpha'\rightarrow 0$, with energies scaled accordingly) focuses on the region close to the branes (small r). Here the "1" in the $H(r)$ drops out, leaving just the $AdS_{5}\times S^{5}$ geometry explicitly 
\begin{equation}
ds^{2}=\frac{r^{2}}{L^{2}}(-dt^{2}+d\vec{x}_{3}^{2})+\frac{L^{2}}{r^{2}}dr^{2}+L^{2}d\Omega_{5}^{2}
\end{equation}
which is exactly the target space in the Maldacena limit. Before taking the limit, the D-branes are explicitly present at a finite location (at $r=0$). After taking the limit, the Maldacena limit zooms infinitely close to the branes. The original position of the branes moves to the boundary at infinity of AdS. After the limit, the explicit branes vanish from the new geometry leaving behind only the curved AdS geometry they created. The gauge theory originally defined by open strings attached on the branes now lives exactly on this boundary at infinity. Thus after the Maldacena limit, you no longer see explicit D-branes in the bulk. They appear only as boundary conditions at infinity encoding the gauge theory on the AdS boundary. The fundamental starting point of string theory is that dynamics of fields on the 2D worldsheet of the string are equivalent to strings propagating through a target space geometry. From the worldsheet perspective we have a 2-dimensional CFT. The fields on this CFT are interpreted as embedding functions $X^{\mu}(\sigma, \tau)$ describing the string position in target space. From a background perspective the target space geometry itself (metric, fields, curvature) appears as coupling constants in this 2D CFT. The fundamental duality at the heart of string theory is the duality between 2D conformal theory on the worldsheet and the target space geometry and fields. Now, specialising to AdS/CFT, in the original setup, we had open strings attached to D-branes which created gauge theory dynamics. From the worldsheet viewpoint, open string endpoints are restricted by boundary conditions given by D-branes. The worldsheet CFT describes open-string modes whose low energy effective theory is gauge theory. From the target space point of view before taking the limit, the closed string geometry curved by D-branes approaches AdS near the branes. In the AdS/CFT limit, the near horizon region transforms into the AdS geometry. Open string gauge theory lives at the boundary and closed strings propagate in the bulk AdS geometry. Therefore the AdS/CFT emerges from a specialised case of the general background worldsheet duality, namely that worldsheet CFT for open strings on D-branes results in a boundary gauge theory (open string sector) whereas closed string worldsheet CFT in curved geometry results in a gravitational AdS bulk (closed-string sector). Thus in AdS/CFT the general background worldsheet correspondence splits into open and closed sectors becoming precisely the open-closed duality. Open strings attached to D-branes result in the boundary gauge theory whereas closed strings in the curved geometry produced by D-branes becomes the bulk gravitational theory. In string theory the term "closed string geometry" generally means the spacetime geometry (target space) in which closed strings propagate. This geometry includes a metric $g_{\mu\nu}(X)$ and fields like the antisymmetric $B_{\mu\nu}$ field, the dilaton $\phi(x)$ and other possible fields. These fields and the metric are backgrounds that appear in the worldsheet sigma model describing closed string propagation. Closed string geometry is dynamically determined from the string theory equations of motion in the form of consistency conditions. Starting with a sigma model describing strings moving in a general target space geometry we have 
\begin{equation}
S_{worldsheet}=\frac{1}{4\pi\alpha'}\int d^{2}\sigma \sqrt{h}h^{ab}g_{\mu\nu}(X)\partial_{a}X^{\mu}\partial_{b}X^{\nu}+...
\end{equation}
Quantum consistency (conformal invariance at the quantum level) of this sigma model imposes conditions on the allowed target space geometries. These conditions become Einstein-type equations (supergravity equations) at low energy, typically 
\begin{equation}
R_{\mu\nu}+2\nabla_{\mu}\nabla_{\nu}\phi-\frac{1}{4}H_{\mu\rho\sigma}H_{\nu}^{\;\;\rho\sigma}+...=0
\end{equation}
The solutions of these equations define consistent closed string backgrounds. The worldsheet-target space duality (often called just T-duality, although this is of course just one special case for the real worldsheet-target duality) refers to a symmetry in string theory that tells us that transformations of the geometry of the target spacetime leave the physics described by the worldsheet theory completely invariant. This duality is fundamental because it reveals an equivalence between seemingly different spacetime backgrounds when viewed from the perspective of the 2-dimensional string worldsheet. As mentioned earlier, in string theory we have two distinct perspectives. On the one side we have the worldsheet viewpoint in which the fundamental theory is a 2-dimensional quantum conformal field theory living on the worldsheet. This CFT describes fields $X^{\mu}(\sigma,\tau)$ which represent the embedding coordinates of the string in some higher dimensional target spacetime. On the other side we have the target space viewpoint. The spacetime geometry emerges as the background fields (metric $g_{\mu\nu}(X)$, antisymmetric tensor field $B_{\mu\nu}(X)$, dilaton $\phi(X)$, etc.) in the sigma model action of the worldsheet theory
\begin{equation}
S=\frac{1}{4\pi\alpha'}\int d^{2}\sigma \sqrt{h}[h^{ab}g_{\mu\nu}(X)\partial_{a}X^{\mu}\partial_{b}X^{\nu}+i\epsilon^{ab}B_{\mu\nu}(X)\partial_{a}X^{\mu}\partial_{b}X^{\nu}+\alpha'R^{(2)}\phi(X)]
\end{equation}
Worldsheet-target space duality means that two apparently distinct sets of target space fields $(g_{\mu\nu},B_{\mu\nu},\phi)$ can correspond to exactly the same physics on the string worldsheet. Thus the physics is invariant under certain transformations of the target space geometry. This duality emerges from the following mathematical property of 2-dimensional quantum field theories. First, the conformal invariance and the modular invariance. String worldsheet theories must be conformally invariant (scale-invariant and invariant under conformal transformations), ensuring consistent quantisation. This invariance strongly restricts possible backgrounds. Second, Buscher rules and gauging sigma-model isometries. If we consider a sigma model with an isometry (a symmetry of the target space metric, like translation symmetry in one direction) then we can gauge this symmetry by introducing gauge fields on the worldsheet, and then integrate out these gauge fields to get a new sigma model which looks different (has new metric and B-field), but because this process is exact at the quantum worldsheet level, the original and the new sigma models are physically equivalent. These transformations are called Buscher transformations and they precisely implement what we call T-duality. As an explicit example, a string moving on a circle of radius R is physically identical to a string moving on a circle of radius $\frac{\alpha'}{R}$. This is the simplest explicit realisation of worldsheet-target space duality. Two distinct target spaces : circles with radius R and radius $\frac{\alpha'}{R}$ are physically identical from the viewpoint of string theory. 
Such a duality exists because strings, unlike particles, have finite extensions and thus probe geometry differently. A particle probes only points. But strings probe extended regions. Thus, two different geometries (e.g. small and large circles) can appear as identical if viewed from a string's perspective, due to the interplay between momentum and winding modes. The momentum modes correspond to standard kinetic energy, sensitive to the large scale structure, while winding modes correspond to strings wrapping around compact dimensions, sensitive to small scale structure. Under the duality transformation, momentum modes in one geometry become winding modes in the dual geometry and vice-versa. This interchange preserves the full physical spectrum, ensuring equivalence. Thus the duality exists because the physics of extended objects (strings) naturally allows for a hidden equivalence between large and small scale geometries. 
However, the worldsheet target space duality includes more structures than simply T-duality. In fact we can have Mirror symmetry, relating two different Calabi-Yau manifolds. Completely different geometric structures give rise to identical worldsheet theories and hence equivalent string theories. Non-Abelian and Poisson-Lie dualities arise as non-Abelian isometries. These include Poisson-Lie dualities as well as generalised geometry transformations. Even holographic dualities like AdS/CFT can be seen as a generalised worldsheet-target space duality since the same worldsheet theory can correspond to a bulk AdS geometry or a boundary gauge theory. Therefore worldsheet target space duality generalises and underlines the idea that spacetime geometry is itself emergent from a deeper quantum theory. This aspect of spacetime emergence is what we will discuss further on. 
\section{emergence} 
It has been shown that D-branes are not simply independently standing higher dimensional objects, but that in fact they emerge as coherent states of strings. In general coherent superpositions are regarded as superpositions of an infinite number of strings. Practically we expect only a finite number of strings to form coherent states, but that would imply a different condition of coherence. Usually a coherent state $\ket{\alpha}$ is defined as an eigenstate of the annihilation operator $a$
\begin{equation}
a\ket{\alpha}=\alpha\ket{\alpha}
\end{equation}
The state $\ket{\alpha}$ can be expanded in terms of number eigenstates $\ket{n}$
\begin{equation}
\ket{\alpha}=e^{-|\alpha|^{2}/2}\sum_{n=0}^{\infty}\frac{\alpha^{n}}{\sqrt{n!}}\ket{n}
\end{equation}
but in this case a coherent state is indeed a superposition of an infinite number of eigenstates $\ket{n}$. When applying this in string theory, in describing D-branes as coherent states of closed strings, it implies a superposition of infinitely many closed string modes. A D-brane is thus viewed as a macroscopic object formed by coherently exciting an infinite number of closed string modes. Such an infinite superposition would ensure several properties. First, it would be an exact eigenstate of the annihilation operator, making it a simple algebraical construction. It would also have a classical like behaviour, namely the coherent states have minimal quantum uncertainty, behaving classically in many aspects. It would also form states that are dynamically stable. Practically however, an infinite superposition can be problematic. To model finite, physical states or to approximate coherent states we will always need some finite truncations. By constructing them, some exact eigenstate properties will be lost, and only approximate coherence will be maintained. Instead of demanding eigenstates of annihilation operators we could impose some alternative phase coherence condition using only finite superpositions. For such truncated coherent states we simply take the coherent state definition and truncate at finite $N$
\begin{equation}
\ket{\alpha}_{N}=\frac{1}{\sqrt{\mathcal{N}}}\sum_{n=0}^{\infty}\frac{\alpha^{n}}{\sqrt{n!}}\ket{n}
\end{equation}
where $\mathcal{N}$ normalises the finite state. Such states are no longer exact eigenstates of $a$ but are close approximations for sufficiently large $N$. Instead of enforcing exact eigenstate coherence, we could impose a weaker condition, namely the phase coherence among modes. Finite superpositions of string states could then be chose carefully such that all modes share a well defined relative phase. Such states would exhibit coherent interference effects without strictly being eigenstates of annihilation operators. Concretely we would define 
\begin{equation}
\ket{\psi_{phase\;coherent}}=\sum_{n=1}^{M}c_{n}e^{i\theta_{n}}\ket{\theta}
\end{equation}
where $\theta_{n}$ are chosen to maximise constructive interference effects. In this case we have a finite superposition, avoiding infinite state complexity, we gain control of coherence through phase alignment rather than an eigenstate condition, and we approximate classical coherent behaviour without needing infinitely many states. Physically, real D-branes should be large, finite objects. Idealising them as infinite coherent superpositions is a theoretical simplification. A finite collection of strings with carefully adjusted relative phases could behave effectively like a coherent macroscopic state (hence like a D-brane). However, such states would be more sensitive to perturbation, losing coherence more easily. In this picture however, D-brane decay would be described in a more realistic sense, leading to insights into the string origin of the Higgs mechanism etc. This point of view resonates with some intuitions we gained from holography and black hole microstates. Black hole microstate constructions in string theory (like fuzzballs) often consider finite combinations of states rather than idealised infinite coherent superpositions. If we can precisely adjust the phases and correlations among finite microstates we can obtain classical geometry like behaviour in holography. Imposing explicit phase coherence, without strict eigenstate conditions would achieve classical behaviour without infinite superpositions and would be relevant in the field of classical geometry emergence. 
If we repeated the above prescription of introducing the Maldacena limit and obtaining the AdS/CFT duality, but starting with D-branes seen as finite superpositions of strings, then we would automatically introduce semiclassical (or slightly quantum) modifications that would remain visible after taking the Maldacena limit. In this way we would obtain a generalised AdS/CFT duality that would contain quantum structure in the AdS region. 
Traditionally, a D-brane is represented as a coherent state of infinitely many closed-string modes 
\begin{equation}
\ket{D}_{standard}=e^{\sum_{n}\alpha_{n}a^{\dagger}_{n}}\ket{0}
\end{equation}
This state satisfies the exact eigenstate conditions $a_{n}\ket{D}=\alpha\ket{D}$. This leads to classical geometry emerging naturally, but the price to pay is infinite superposition, making the description unrealistic. If on the other hand we would generate an explicitly finite superposition of closed state modes
\begin{equation}
\ket{D}_{finite}-\sum_{n=1}^{N}c_{n}e^{i\theta_{n}}\ket{n}
\end{equation}
with $N$ finite, $c_{n}$ are real coefficients determining the amplitude, and $\theta_{n}$ are carefully chosen phases ensuring phase coherence among string modes, we would obtain a state that is not an eigenstate of any annihilation operator $a_{n}$ but that would retain phase coherence. All phases $\theta_{n}$ are chosen to maximise constructive interference for certain observables, such as the position or tension of the brane. We could choose $\theta_{n}=n\phi$ for some fixed $\phi$, ensuring constructive interference at particular points in target space. The explicit state would be 
\begin{equation}
\ket{D}_{finite}=\frac{1}{\sqrt{N}}\sum_{n=1}^{N}e^{i n \phi}\ket{n}
\end{equation}
This finite state is not an eigenstate of any annihilation operator, but the carefully chosen phases mean that for particular observables (like the energy distribution, or position distribution) we can get strongly peaked coherent interference effects resembling a macroscopical classical object, we would have a "quantum brane" that is very close to a classical brane in certain aspects. The D-brane indeed retains fundamental quantum features. Unlike ideal coherent states, this finite phase coherent state does not saturate minimal uncertainty exactly. It instead retains finite uncertainty in position, momentum, or tension. It also retains fluctuations in geometry and fields around the brane. Thus, such a finite quantum brane naturally encodes quantum gravity like fluctuations and uncertainty in its geometry. If phases are disturbed (e.g. due to interactions), coherence may diminish and thus the brane is explicitly a quantum coherent object. In such a model quantum decohernece can be directly studied. 
Since this brane is finite and not an exact eigenstate, it can naturally decay, change, or re-arrange quantum mechanically. Thus, such "finite branes" explicitly carry quantum instability features, and their lifetimes become calculable quantum mechanical observables. We would obtain a finite dimensional Hilbert space which directly simplifies the analysis of entanglement and other quantum processes involving D-branes. Moreover, the finite superposition explicitly encodes quantum gravitational fluctuations and can produce predictions for observable quantum gravity signatures. Now, having defined this finite superposition leading to practical finite coherent behaviour, we could ask what exactly would mean to have a worldsheet-target space dualtiy (or in particular T-duality) in this semiclassical limit, when the target geometry itself is represented by coherent states or finite coherent superpositions of string modes. Usually T-duality is understood by starting from strings propagating in some geometry (for example a circle of radius R), and then T-duality is applied to transform this duality into a dual geometry, (for example a circle of radius $\alpha'/{R}$), exchanging momentum modes and winding modes. At the quantum (exact) worldsheet level, the physics of both descriptions is identical. The two seemingly distinct spacetime backgrounds (target spaces) are physically indistinguishable. Consider however now that the target space geometry itself is represented not as a fixed classical geometry, but rather by a semiclassical coherent superposition of string modes. For example a target space geometry like a circle of radius R, or even a D-brane, is explicitly built as a coherent or finite coherent superposition of closed string excitations 
\begin{equation}
\ket{\psi(R)}=\sum_{n=1}^{N}c_{n}(R)e^{i\theta_{n}(R)}\ket{n}
\end{equation}
Here the state $\ket{\psi(R)}$ encodes the geometry semiclassically through coherent interference. The geometry emerges approximately as classical from interference patterns of finite quantum states. In this coherent superposition scenario, T-duality now becomes an equivalence of coherent superposition states. A coherent superposition representing a large geometry must map explicitly to another coherent superposition representing a small geometry with momentum and winding modes exchanged
\begin{equation}
\ket{\psi(R)}\leftrightarrow \ket{\tilde{\psi}(\alpha'/R)}
\end{equation}
The dual state $\ket{\tilde{\psi}(\alpha'/R)}$ encodes the dual geometry via a different choice of mode amplitudes and phases which now emphasise winding instead of momentum. Thus, the duality relates two coherent quantum states, the original coherent superposition of radius R, dominated by momentum modes, and the dual coherent superposition of radius $\frac{\alpha'}{R}$ dominated by winding modes. This quantum state viewpoint clarifies how geometry emerges from quantum superpositions and how dualities map these quantum coherent states into each other. The original state $\ket{\psi(R)}$ is a semiclassical coherent interference that creates a geometry that is approximately classical at radius $R$. The dual state $\ket{\tilde{\psi}(\alpha'/R)}$ represents a quantum coherence that now rearranges the modes so that interference patterns yield approximately classical geometry at radius $\alpha'/R$. Therefore T-duality is a change of basis in quantum interference space, mapping momentum mode coherence to winding mode coherence. The geometry is fundamentally emergent from quantum coherence rather than a classical geometry. The semiclassical coherent superpositions encode quantum uncertainty about geometry. The duality explicitly transforms how uncertainty manifests. Momentum uncertainty transforms to winding uncertainty. The duality is therefore a quantum equivalence of coherent states rather than of classical backgrounds. In this case therefore we explicitly clarify the quantum gravity interpretation of the duality, showing how duality arises naturally from coherent superpositions and quantum interference patterns. 
This approach modifies also what we mean when we say "a string is moving through spacetime". In reality a string would be more excited in a sea of coherently superposed strings that form a classical or semiclassical background state which would then be associated to spacetime. However, this will have an impact on how we understand the idea of distance on a string and how we would understand the concept of "string length". 
The usual intuitive image claiming that "a string is a one-dimensional object moving through a fixed, classical spacetime" cannot be fundamentally accurate in this new fully quantum (or at least semiclassical) picture. Instead, spacetime itself is a coherent superposition (a quantum interference pattern) of a large number of string excitations. A string moving in spacetime is not really a distinct object travelling independently through a fixed background but rather corresponds to changes in excitation and interference within a large quantum superposition state. The string is not literally moving through spacetime. Rather, it corresponds to an excitation or disturbance in a quantum sea of coherently superposed string modes that collectively form spacetime itself. If however spacetime itself is an emergent quantum interference pattern, then the classical notion of distance or length fundamentally depends on having a pre-existing geometry. If spacetime is emergent there is no fundamental geometry to start with. The geometry itself emerges from quantum correlations and interference. Therefore distance on a string is not fundamental but emergent. The idea of the length of a string is not fundamentally well-defined. Rather, the string's length must also emerge from quantum interference patterns. This is consistent with the modern viewpoint of emergent geometry in quantum gravity. If geometry including length and distance emerges from quantum coherence, then string length itself is an approximate emergent quantity defined by expectation values and correlations among quantum states. Concretely, we would define length through quantum expectation values of suitably defined observables. For example, if we consider a string state $\ket{\Psi}$, its emergent length $L$ would be defined something like 
\begin{equation}
L=\bra{\Psi}\hat{L}\ket{\Psi}
\end{equation}
where $\hat{L}$ is a suitable quantum operator representing the emergent notion of length. But the operator $\hat{L}$ doesn't come from a classical metric, since the metric itself emerges. Rather $\hat{L}$ must be constructed from correlation functions or quantum coherence measures. We could define distance or length via quantum correlation functions between different parts of the superposition. Quantum entanglement or correlation structure among modes naturally gives rise to an emergent metric structure. Thus the length of the string emerges explicitly as a quantum correlation based observable, not a classical geometric one. Consider a simplified toy model. Suppose the string is described by a finite coherent superposition
\begin{equation}
\ket{\Psi}=\sum_{n=1}^{N}c_{n}e^{i\theta_{n}}\ket{n}
\end{equation}
This would imply that the string is an "emanation" of the D-brane seen as a coherent superposition of strings itself. The individual open string is therefore an excited state of such a semiclassical D-brane. 
Define a length operator as an operator sensitive to interference between these modes. For example 
\begin{equation}
\hat{L}\sim\sum_{n,m}L_{n,m}\ket{n}\bra{m}
\end{equation}
The matrix elements $L_{m,n}$ encode correlation-based distances emerging from quantum coherence structure. We can pick diagonal terms (classical approximation of length) or off-diagonal terms (that would encode quantum interference corrections to length). Then the length of the string emerges as 
\begin{equation}
L=\bra{\Psi}\hat{L}\ket{\Psi}=\sum_{n,m}c_{n}c_{m}^{*}e^{i(\theta_{n}-\theta_{m})}L_{n,m}
\end{equation}
This shows how the length is quantum coherence dependent, emergent and subject to quantum uncertainty. Therefore no fundamental geometric distances exists. Geometry, and thus length or distances, emerges entirely from quantum coherence and correlation. Quantum uncertainty is introduced in geometry. The fluctuations of length and geometry become explicit and fundamental, as we would expect from quantum gravity. We therefore notice that strings do not literally move through space. Instead they represent quantum excitations in an emergent, coherent superposition of string states, forming spacetime itself. The very notion of length or distance on a string is not fundamental, but instead emerges as a quantum observable defined by quantum coherence and interference. 
One may ask what exactly is "interfering" if the string doesn't have a length. This question is very similar to the question as of how the photon encodes wavelength when an individual photon has no spatial extension but only energy. We may therefore ask how precisely the length of a string is encoded when we talk about a string in the usual sense. 
\section{wavelength from a dimensionless photon, and emerging length from an algebraic string}
A single photon doesn't have an explicit spatial length. It is described by a quantum state with energy $E=h\nu=\frac{hc}{\lambda}$, and its "length" (wavelength) emerges from its quantum state, and is encoded in the frequency (energy). Thus the photon's "length" (the wavelength) emerges not from spatial extension, but from quantum interference properties that determine its frequency or energy. Similarly, a string doesn't explicitly have a classical spatial length yet defined. Instead its "length" emerges from quantum interference between different string modes, analogous to photon wavelength emerging from energy interference. 
We start by constructing explicitly a quantum length operator $\hat{L}$. We describe a string quantum state as a coherent or finite coherent superposition of modes 
\begin{equation}
\ket{\Psi}=\sum_{n}c_{n}^{i\theta_{n}}\ket{n}
\end{equation}
These modes $\ket{n}$ represent different string mode excitations, not yet spatially extended objects. The length operator must be defined via the correlation among these quantum modes 
\begin{equation}
\hat{L}=\sum_{n,m}\ket{n}\bra{m}
\end{equation}
The matrix elements $L_{n,m}$ encode the notion of length as correlations between modes. The diagonal terms represent intrinsic length associated with each individual mode, while the off diagonal terms encode quantum interference effects between different modes. 
For example we could chose a simple form 
\begin{equation}
L_{n,m}=L_{0}\delta_{n,m}+le^{-\frac{(n-m)^{2}}{2\sigma^{2}}}
\end{equation}
$L_{0}$ represents a classical baseline length, l is the interference length scale, controlling the strength of quantum interference, $\sigma$ determines the range of quantum coherence or interference. This defines length via quantum coherence structure (interference between modeS). Using this operator we cal calculate the string length expectation value for the finite quantum superposition state
\begin{equation}
L=\bra{\Psi}\hat{L}\ket{\Psi}=\sum_{n,m}c_{n}c_{m}^{*}e^{i(\theta_{n}-\theta_{m})}L_{n,m}
\end{equation}
This shows how length emerges from two types of contributions explicitly. The classical contribution, for diagonal terms (n=m) giving a classical approximation 
\begin{equation}
L_{classical}=\sum_{n}|c_{n}|^{2}L_{nn}
\end{equation}
and the quantum interference terms, the off diagonal terms ($n\neq m$), explicitly encoding quantum coherence corrections
\begin{equation}
L_{quantum\;\;interference}=\sum_{n\neq m}c_{n}c_{m}^{*}e^{i(\theta_n-\theta_m)}L_{n,m}
\end{equation}
The interference terms encode quantum corrections or fluctuations to the classical length arising from coherence structure among modes. But what exactly is interfering if the string doesn't have a length yet? Initially there is no classical string length. What is interfering are the amplitudes and phase of different quantum string modes. Each string mode, like a photon frequency mode, encodes a quantum number (analogous to photon frequency or energy). The interference patterns among these modes encoded in their phases $\theta_{n}$ gives rise to a coherent spatial pattern, thus defining the notion of length. The interference is in the space of mode excitations and their quantum numbers, not explicitly in a pre-exiting spatial geometry. Similar to a photon, what interferes are not spatially extended objects but quantum states distinguished explicitly by energy/mode numbers, phases, and quantum numbers. Thus the string length emerges not from spatial extension but from interference in mode space, analogous to electromagnetic wavelengths emerging from frequency/energy space interference. When speaking of a "real unique string" its length doesn't come from spatial geometry explicitly but from quantum coherence patterns. The string length is encoded entirely in quantum coherence between modes. Measuring a real string state means observing quantum interference patterns which define the length of the string. Just as the photon's wavelength is not about spatial size, but energy interference, a string's length is encoded in the amplitude phase interference among its quantum modes. 
\section{quantum numbers and spacetime symmetries}
Quantum numbers are essentially labels of distinct states of a quantum system. They arise naturally whenever we have a quantum system defined by some underlying symmetry or structure. Photons have quantum numbers like frequency (or equivalently energy) and momentum, due to the underlying symmetry of spacetime translation invariance. Strings, similarly have quantum numbers such as excitation levels, momentum, winding number, oscillator numbers, and mode frequencies, arising from the symmetry and structure of the worldsheet conformal field theory. These quantum numbers are fundamentally labels that organise the possible states the system can occupy. They emerge naturally from solving the underlying equations (wave equations for photons, conformal invariance conditions for strings). Quantum numbers emerge from the underlying equations of motion and symmetries of the theory. Photons arise as solutions of the quantum electromagnetic wave equation:
\begin{equation}
\begin{array}{cc}
\hat{H}\ket{E}=E\ket{E},& \hat{P}\ket{p}=p\ket{p}
\end{array}
\end{equation}
Here, $\hat{H}$ and $\hat{P}$ are the Hamiltonian (energy) and momentum operators. Energy and momentum thus appear as eigenvalues labeling quantum states. 
Strings similarly arise as quantum solutions of the two-dimensional conformal field theory describing the worldsheet
\begin{equation}
(L_{0}-a)\ket{string\;mode}=0
\end{equation}
where the Virasoro operators $L_{0}$ encode mode frequencies, energies, oscillator numbers, etc. Thus string quantum numbers (oscillator excitation levels, momenta, winding numbers) arise naturally from conformal invariance constraints on the worldsheet. In other words, quantum numbers emerge because quantum states must satisfy particular equations (wave equations or conformal constraints) and these equations naturally yield discrete (or continuous) quantum labels that we call "quantum numbers".
Quantum numbers by themselves do not explicitly refer to spatial geometry or length. A photon's quantum number "frequency" doesn't directly imply spatial size, but it implicitly defines a length scale (the wavelength) through the relation 
\begin{equation}
\lambda=\frac{c}{\nu}=\frac{hc}{E}
\end{equation}
Similarly string quantum numbers do not explicitly imply length by themselves. Instead, the length emerges implicitly as an interference pattern or correlation among different quantum number labeled states. Quantum numbers label distinct states and the geometry emerges from the relationship and interference between these states, not from individual quantum numbers alone. For a photon, the wavelength is not stored explicitly spatially. Rather wavelength emerges from interference of multiple photon frequency states (Fourier modes). Thus wavelength is fundamentally defined through frequency-space correlations and interference patterns. In string theory, a string's length is not stored explicitly in a spatial extension. Rather it emerges implicitly from quantum coherence (interference) among multiple string states labeled by quantum numbers. The interference among these quantum states yields a spatial pattern or correlation which we identify as length and thus geometry. Consider two string modes labeled by quantum numbers n and m. Each mode individually has no explicit spatial geometry, only quantum numbers. Now, build a quantum state involveing these two modes 
\begin{equation}
\ket{\Psi}=c_{n}e^{i\theta_n}\ket{n}+c_{m}e^{i\theta_m}\ket{m}
\end{equation}
The length emerges only if we measure an observable sensitive to interference between these modes. For instance consider the expectation of our length operator $\hat{L}$
\begin{equation}
L=\bra{\Psi}\hat{L}\ket{\Psi}=|c_{n}|^{2}L_{nn}+|c_{m}|^{2}L_{mm}+c_{n}c_{m}^{*}e^{i(\theta_n-\theta_{m})}L_{nm}+c_{m}c_{n}^{*}e^{-i(\theta_n-\theta_m)}L_{mn}
\end{equation}
The off diagonal interference terms create a correlation structure between modes labelled by different quantum numbers. This correlation structure is precisely what defines an emergent length. Without interference terms, length would simply be an average of diagonal classical lengths. The quantum interference explicitly modifies and defines the emergent geometry. Thus length emerges not directly from quantum numbers, but rather from correlations and coherence between different states labeled by those quantum numbers. In the semiclassical limit, many quantum states interfere coherently, and the resulting interference pattern becomes sharply defined, stable, and classical like. The correlation among these quantum number labeled states becomes stable and produce a coherent macroscopic pattern. This stable interference pattern can be interpreted as classical geometry, distances, and metrics. Thus, a classical length and metric are simply the stable macroscopic interference patterns produced by coherent superpositions of quantum number labeled states. 
However, quantum numbers arise whenever we have a quantum system defined by some underlying symmetry or structure. In particular for photons, quantum numbers like frequency or energy and momentum appear due to the underlying symmetry of spacetime translation invariance. If we want to maintain the model of spacetime and geometry emergence only from coherent superpositions of modes characterised by these quantum numbers when taking a semiclassical limit, we may ask whether we assume the existence of a spacetime with translation symmetry? It seems like a circular argument appeared. Quantum numbers (such as energy or momentum quantum numbers) are usually defined using symmetries of spacetime, for example momentum arises from translation invariance. But spacetime itself emerges from quantum coherence and interference of states labeled by these quantum numbers. However, this circularity is only apparent. In quantum field theory, quantum fields (photons, strings, particles) are defined on a fixed, pre-existing classical spacetime. The symmetries of this spacetime (translation, rotation, Lorentz invariance) directly define the quantum numbers labelling sates. The quantum gravity perspective is that spacetime is not fundamental but instead it emerges from quantum coherence and correlation structures among quantum states. Thus, the spacetime symmetries themselves must also emerge. The apparent circularity arises because we use reasoning from both perspectives, simultaneously without separating them. 
Fundamentally, spacetime geometry and its symmetries do not exist yet. Instead, there are only abstract quantum states, not yet labeled by classical geometric symmetries. Quantum states at this fundamental stage do not have quantum numbers defined by spacetime symmetries explicitly. Quantum numbers initially arise purely from the algebraic structure or internal symmetries of the fundamental quantum system (for instance, conformal symmetry on the string worldsheet or internal algebraic symmetries in abstract quantum gravity frameworks). Then at the level of large coherent superpositions of states, stable correlations (coherence patterns) emerge. These stable coherence patterns can be approximately interpreted as having translation invariance or spacetime symmetries in an emergent classical sense. Only after this emergent structure stabilises do we retroactively interpret certain quantum labels as momentum, energy, frequency, related to translation symmetry. Initially, these labels were abstract quantum numbers unrelated explicitly to spacetime geometry. Thus the important step in breaking the apparent circularity is to realise that quantum numbers initially exist without reference to spacetime geometry, arising purely from the quantum system's abstract algebraic structure. Spacetime geometry (and thus classical symmetries) emerges later as a stable coherent approximation. Once this geometry emerges approximately, we retroactively identify certain quantum numbers as being associated with translation, rotation, or Lorentz invariances. Therefore quantum numbers were initially algebraic (non-geometric). They only later become geometric (momentum, energy) after spacetime geometry emerges. 
For example, initially the fundamental quantum states are described by abstract labels $\ket{a}, \ket{b}, \ket{c},...$ arising from internal symmetries or algebraic structures unrelated to spacetime geometry. Coherent superpositions of these abstract states form stable interference patterns. Some stable interference patterns behave like plane wave states
\begin{equation}
\ket{\psi}=\sum_{a}c_{a}e^{i\alpha_a}\ket{a}
\end{equation}
These coherent patterns approximate something like plane waves, which in classical physics carry definite momentum. Once these stable plane wave like coherence patterns emerge, we retroactively identify the labels $a$ as momentum quantum numbers, because the emergent structure looks like translation invariance. But initially, these labels were purely algebraic, not explicitly geometric. 
In string theory initially, quantum numbers come from internal worldsheet symmetries (conformal algebra). These symmetries exist independently of target space geometry. Large coherent superpositions of string states form stable interference patterns. At large scales, these interference patterns define an emergent spacetime geometry. Once spacetime geometry emerges approximately, certain worldsheet quantum numbers (oscillator numbers, mode labels, etc) can be reinterpreted as spacetime momentum or energy. Initially they were purely algebraic labels on the string worldsheet. 
Therefore the chain of reasoning is : Abstract quantum numbers $\rightarrow$ coherent interference patterns $\rightarrow$ emergent approximate geometry $\rightarrow$ geometric interpretation of quantum numbers. 
\section{Target space}
Following the observations above, we notice that initially we have no classical target space. We only have an abstract worldsheet conformal field theory that describes the quantum modes of a string. These quantum modes are labelled by purely abstract numbers arising from internal symmetries and algebraic structure of the worldsheet theory (conformal invariance, Virasoro algebra, mode expansion). When a large number of these abstract worldsheet modes enter into coherent superpositions, they produce stable interference patterns. These stable interference patterns are precisely what we interpret as emergent classical target space (background geometry) in string theory. Thus the target space geometry (spacetime) emerges purely from stable coherent patterns among abstract internal modes defined initially only on the worldsheet. A traditional worldsheet-target space duality or correspondence states that a 2D worldsheet theory of strings is exactly equivalent to strings moving in a particular target space geometry. In order words, certain worldsheet CFTs directly encode specific classical target space geometries. Initially there is only the worldsheet CFT, with abstract quantum modes. Large coherent superpositions of these worldsheet modes create stable quantum interference patterns. These stable interference patterns become the classical geometry of the target space. Thus what looks from a target space viewpoint as a classical geometry is nothing more than coherent quantum patterns formed by worldsheet modes. Therefore the worldsheet $\rightarrow$ target space correspondence is precisely the statement that coherent interference patterns of worldsheet modes are the target space geometry. In a sense the worldsheet/target space correspondence is reduced to the problem of taking a semiclassical limit and it relates to the problem of obtaining classical physics from quantum mechanics. Consider a simplified illustrative example. From the worldsheet perspective consider modes labeled abstractly by integers $n$. The quantum state is 
\begin{equation}
\ket{\Psi}=\sum_{n}c_{n}e^{i\theta_{n}}\ket{n}
\end{equation}
From the target space perspective, the stable coherence among these abstract modes forms stable patterns. If these stable coherence patterns appear like plane-wave superpositions with well defined frequencies and wavelengths, we interpret this emergent stable coherence as classical spacetime geometry. The large stable coherence is the emergent classical geometry. The plane wave like coherence s translation invariance, and the translation invariance defines momentum, energy, and the metric. This viewpoint clarifies other dualiities as well. For example T-duality (changing radius $R\leftrightarrow\frac{\alpha'}{R}$) corresponds to changing the quantum interference patterns and coherence between modes. Thus the emergent classical geometry changes from large to small circles purely through changing quantum coherence structure. Mirror symmetry (Calabi-Yau spaces) appears because different coherence patterns (different superpositions of worldsheet modes) yield distinct classical geometries. Mirror symmetry explicitly states that distinct quantum coherence patterns can give rise to two different classicla geometries yet produce identical worldsheet physics. 
Spacetime symmetry and geometry arise as stable emergent properties from large scale quantum coherence and quantum numbers are initially abstract algebraic labels, and only become geometric once stable geometry emerges. 
However one may ask how can we talk about quantum numbers, worldsheet, or internal algebra without implicitly assuming geometry? An algebraic quantum number is a label assigned to a quantum state solely based on the structure of the algebra of operators that define the quantum theory, without reference to geometry. For example, consider the standard algebraic structure of quantum mechanics. We have a set of operators forming an algebra (for example, creation and annihilation operators $a$ and $a^{\dagger}$, or symmetry operators $J^{2}$, $J_{z}$, or Virasoro operators $L_{n}$). 
Quantum numbers are simply eigenvalues or labels assigned to states based on these algebraic operators 
\begin{equation}
\begin{array}{cc}
L_{0}\ket{n}=n\ket{n},&J_{z}\ket{m}=m\ket{m}
\end{array}
\end{equation}
Initially these quantum numbers are purely algebraic in the sense that they don't rely on geometric notions like position or length. They arise solely from the algebraic consistency conditions of the theory (such as closure, commutation relations, eigenvalue equations). Spin quantum numbers for example $m=-j,...,j$ are purely algebraic labels defined by the algebra of angular momentum operators, without ever requiring explicit geometry. 
In standard string theory the worldsheet is typically described geometrically as a two-dimensional surface swept out by a one-dimensional string in spacetime. However, at the most fundamental level, such as in a fully quantum theory of strings, the worldsheet need not be geometric in an explicit classical sense. Instead, the worldsheet can be viewed purely algebraically as a conformal field theory. It is not a geometric object, rather, it is a quantum theory described purely by an algebra of local operators, conformal symmetry generators, and their correlation functions. The Virasoro algebra generated by operators $L_n$ completely characterises this "worldsheet" theory algebraically, without explicit geometry. 
Thus the worldsheet is nothing more than the algebraic structure of a conformal field theory. It only acquires geometric meaning at a higher, emergent, semiclassical level. 
We can indeed define a worldsheet in a completely non-geometric way. Begin with an algebraic structure, defined by operator algebras, for example the Virasoro algebra
\begin{equation}
[L_{m},L_{n}]=(m-n)L_{m+n}+\frac{c}{12}(m^{3}-m)\delta_{m+n,0}
\end{equation}
and define the states as algebraic objects that are representations of this algebra 
\begin{equation}
\begin{array}{cc}
L_{0}\ket{h}=h\ket{h}, &L_{n}\ket{h}=0,\;\; (n>0)
\end{array}
\end{equation}
Correlation functions and quantum numbers are fully defined by algebraic consistency conditions, like Ward identities or conformal bootstrap equations. At this stage we have constructed a worldsheet theory without ever introducing geometric notions explicitly. We may ask if internal geometry is even necessary. In fact, at the most fundamental level, it is indeed only internal algebra that is necessary. Geometry is not fundamental but emergent. Initially we have an algebra of abstract operators, and states labeled purely algebraically by quantum numbers which are eigenvalues of these algebraic operators. Internal geometry, such as compact extra dimensions, worldsheet coordinates or target space coordinates is not fundamental but a convenient way to represent algebraic structures at a semiclassical level. Geometry only emerges when we take large coherent superpositions of states labeled algebraically. Then stable interference patterns appear, which can be interpreted classically as geometry. Therefore even internal geometry is never fundamentally required. We can present a very simple analogy, starting for example from the harmonic oscillator algebra
\begin{equation}
\begin{array}{ccc}
[a,a^{\dagger}]=1, & \hat{H}=a^{\dagger}a+\frac{1}{2}, & \hat{H}\ket{n}=(n+\frac{1}{2})\ket{n}
\end{array}
\end{equation}
Initially quantum numbers $n$ label algebraic states purely abstractly. Now considered coherent states, as large superpositions
\begin{equation}
\ket{\alpha}=e^{-|\alpha|^{2}/2}\sum_{n=0}^{N}\frac{\alpha^{n}}{\sqrt{n!}}\ket{n}
\end{equation}
Such coherent superpositions produce stable, classical-like interference patterns. These stable interference patterns resemble classical states with position and momentum that become sharply defined and classical like even though initially there was no explicit position or momentum. 
But how exactly do quantum correlations between abstract algebraic modes translate precisely into the notions we call "time", "space", or "length"? 
Geometry, at its most basic classical level means that there is some notion of distance or separation between points or events. But fundamentally what is distance? Distance and length fundamentally mean that certain events or states are related or correlated in a structured and consistent way. Classical geometry arises precisely when correlations between states become stable, consistent and predictable. Without stable correlations, no meaningful geometry can arise. Thus, at the fundamental quantum level, initially we only have abstract quantum states labelled algebraically. The way in which geometry can appear from purely quantum objects is through stable correlations between these abstract states. When correlations become large scale, stable, and consistent, we begin to interpret them classically as "geometry". Two points separated by small distance are strongly correlated in a geometric sense. Points separated by a large distance have weaker correlations. Thus, strength of correlation encodes distance. When quantum states labeled abstractly become correlated in a stable pattern, such as coherent interference, the pattern itself encodes a notion of distance, defining geometry. Suppose we have quantum states labelled by abstract indices n,m. Define correlation strength between these states. States that are strongly correlated represent close points while weakly correlated states represent distant points 
\begin{equation}
distance(n,m)\sim\frac{1}{correlation\;\;strength\;\;between\;\;\ket{n},\ket{m}}
\end{equation}
Time similarly emerges from correlations but now we focus on correlations involving ordered change or evolution. Time fundamentally means events or states have a consistent sequential ordering, one event reliably follows another. This sequential order emerges if quantum states have stable correlations that evolve systematically under a quantum Hamiltonian or some algebraic operator. Time emerges because correlations between states evolve under the action of a Hamiltonian or algebraic operator 
\begin{equation}
\ket{\psi(t)}=e^{-iHt}\ket{\psi{0}}
\end{equation}
A state $\ket{\psi(t_{2})}$ reliably follows $\ket{\psi(t_{1})}$. Such stable and sequential correlations between states define time. 
Consider a simple algebraic example. Suppose we have an abstract algebra of operators $A_{n}$ labelling states $\ket{n}$. Construct correlation functions 
\begin{equation}
C_{nm}=\bra{n}A_{n}A_{m}\ket{m}
\end{equation}
Strong correlations (large $C_{nm}$) mean points/states are close. Weak correlations mean they are far apart. You can thus define a metric explicitly from correlations 
\begin{equation}
g_{nm}\sim \frac{1}{C_{nm}}
\end{equation}
For the time correlations $\bra{\psi_{n}(t_{1})}\ket{\psi_{m}(t_{2})}$ defines the sequential ordering of states. Stable, ordered correlations yield the emergent notion of time. 
Thus, both space and time explicitly arise purely from the quantum correlation structure. 
Classical geometry emerges as a coarse-grained approximation. Quantum coherence patterns involving enormous numbers of states become extremely stable through decoherence and entanglement. These stable macroscopic patterns are indistinguishable from classical geometric notions. Thus classical geometry is just the macroscopic stable correlation pattern among many quantum states (like how classical waves emerge from coherent interference of enormous numbers of photons). 
The problem of the origin of spacetime therefore becomes the problem of how semiclassical or classical limits emerge from quantum mechanics. However, there are many such classical or semiclassical limits. The reason we usually see only one consistent classical geometry is fundamentally linked to the stability, consistency, and selection of certain interference patterns. While quantum mechanics allows for numerous classical like behaviours, not all are stable or self-consistent. Most quantum interference patterns "wash out", fluctuate chaotically, or decohere rapidly. The observed spacetime emerges precisely from those quantum correlations that maximise stability, coherence and consistency over extremely large scales, both spatial and temporal. Thus the observed classical geometry emerges as the most stable and robust large scale interference pattern selected by the dynamics of quantum coherence and decoherence. Special relativity arises because of the fact that spatial and temporal correlations are not independent. They are connected via quantum coherence structure, forming precisely the Minkowski structure of spacetime. The causal structure (light cones, causality) emerges naturally from how correlations propagate in the quantum state space, constrained by coherence and entanglement structures. The Lorentz symmetry emerges precisely as the symmetry of these quantum correlations. It appears as an approximate symmetry of stable quantum coherence patterns selected by consistency and stability criteria.
\section{closed causal loop}
We could ask what kind of quantum correlations would be necessary to generate an emergent classical geometry that includes closed causal loops? A closed causal loop is a classical geometric construction in which a trajectory loops back upon itself in both space and time. Such loops allow events in the future to causally influence events in their own past. Classically these are typically associated with strong spacetime curvatures or exotic geometries (wormholes, rotating black holes, Goedel Universes). 
However, classical causal structures emerge entirely from stable quantum correlations among abstract states. Therefore to generate something like a closed causal loop, quantum correlations would have to differ significantly from standard correlations that yield normal causal structure. Let us first look at what kind of quantum correlations produce standard causal structure. The standard causality emerges if quantum states are correlated sequentially in a linear and consistent ordering 
\begin{equation}
\ket{\psi(t_{1})}\rightarrow\ket{\psi(t_{2})}\rightarrow \ket{\psi(t_{3})}...
\end{equation}
These correlations are time-ordered forming a clear direction of causality. Correlations among spatial points occur simultaneously (in a relativistic sense) and do not violate linear causal ordering. In order to obtain a closed causal loop, quantum correlations must differ significantly from this standard pattern. We would require non-linear sequential correlations. Instead of a linear ordering, quantum states must exhibit some cyclic correlations
\begin{equation}
\ket{\psi(t_{1})}\rightarrow\ket{\psi(t_{2})}\rightarrow \ket{\psi(t_{3})}...\rightarrow \ket{\psi(t_{n})}\rightarrow\ket{\psi(t_{1})}
\end{equation}
This cyclic correlation means that the state $\ket{\psi(t_{1})}$ would be strongly correlated with itself at a later point in its sequential evolution, producing a causal loop. 
Correlations would exist that link states "forward in time" explicitly back to earlier states in the loop. This implies a very strong coherence or entanglement structure across what we normally consider temporally distant or unrelated states. This kind of correlation is strongly non-local and non-causal in the traditional sense. The interference pattern among quantum states would have to exhibit topologically non-trivial coherent structures forming cyclic or toroidal coherence patterns in the space of quantum states. Normal spacetime correlations correspond to linear or simply connected coherence patterns. Closed causal loops would correspond to coherence patterns explicitly forming closed loops or rings in correlation space. 
Explicitly define a correlation matrix that is cyclic 
\begin{equation}
\begin{array}{cc}
C_{ij}=\Bracket{\psi_{i}|\psi_{j}}, &C_{i,i+1},C_{n,1}\neq 0
\end{array}
\end{equation}
Such a cyclic correlation structure would yield an emergent classical geometry that includes closed timelike loops. However, would these cyclic quantum correlations produce stable, physically realisable geometry?
In normal circumstances we would think that such cyclic coherence patterns would be highly unstable and delicate. Quantum decoherence would tend to rapidly destroy cyclic coherence preventing stable closed loops from persisting. However, it is know, as I stated previously, that taking a classical limit is not a unique process, and different paths of taking the limits offer different types of outcomes. In normal circumstances, quantum decoherence would tend to rapidly destroy cyclic coherence preventing stable closed timelike loops from persisting. 
In fact, there is not just one unique way of taking a semiclassical limit from quantum states. Indeed, quantum theory admits multiple possible semiclassical or classical regimes, depending on how one takes limits (such as coherent states, decoherence mechanisms, choices of observables, etc.)
The particular classical geometry that emerges is highly dependent on exactly how the classical limit is taken. Therefore it is plausible that if we choose a different, possibly more robust or subtle way of taking the semiclassical limit, we might find that certain correlation structures previously dismissed as unstable or fragile, become stable and robust, thus allowing exotic classical geometries such as those with closed causal loops, to naturally emerge. The classical limit is essentially a procedure for extracting stable classical-like behaviour from a fundamentally quantum theory. Different criteria or procedures could yield different classical-like behaviours. For example when using coherent states as before, we emphasise minimal uncertainty and classical like wave-packets. But when we use decoherence, we emphasise how quantum systems interact with their environment, selecting certain states as classical. For the large-N or thermodynamic limits we emphasise collective stable patterns in very large quantum systems. We may also have eigenstate thermalisation or quantum typicality through which we assume that classical limits pick states according to their typicality in high-dimensional Hilbert spaces. Thus, the classical geometry we see depends strongly on which type of limit or criteria we use and how we engineer the phase relations or typicality of the potential quantum states. The standard classical spacetime we experience might just be the geometry emerging naturally under the most common or physically relevant limit, but alternative classical limits could produce very different geometry. For example, in a standard classical limit, quantum coherence rapidly decoheres. Classical geometry emerges simply in a stable and causal form (no closed loops). However, we could engineer a decoherence mechanism and boundary conditions or large scale coherence structures that would allow quantum correlations to stabilise in a different way. The classical geometry would then emerge with exotic structures like stable causal loops. An example would be a topological quantum state that may be obtained in a quantum spin liquid. 
Let us present a comparison using a simplified toy model for a quantum spin liquid. We compare two different semiclassical limits. In limit A we produce an emergent classical Minkowski type causal structure and in Limit B we produce an emergent classical geometry containing closed causal loops. 
Consider a simplified quantum spin liquid system described by spins on a lattice with N sites. The quantum states $\{\ket{s_{i}}\}$ with $i=1,2,...,N$ represents lattice positions. Each state is labelled by spin quantum numbers $(\uparrow,\downarrow)$. A generic quantum state as superposition is 
\begin{equation}
\ket{\Psi}=\sum_{s_{1},...,s_{N}}c_{s_{1},...,s_{N}}\ket{s_{1},...,s_{N}}
\end{equation}
Define a correlation function between two sites $i$ and $j$
\begin{equation}
C_{ij}=\bra{\Psi}\hat{S}_{i}^{z}\hat{S}_{j}^{z}\ket{\Psi}
\end{equation}
where $\hat{S}_{i}^{z}$ is the spin operator in the $z$-direction at site $i$. 
For the case of the semiclassical limit $A$ (a Minkowski causal structure), correlations must be consistent with the standard causal ordering and locality. Correlations decay quickly with spatial distance, and no loops in causal structure occur. We define the coefficients such that the ground state is dominated by short-range correlations (nearest neighbour correlations) 
\begin{equation}
\begin{array}{cc}
c_{s_{1},...,s_{N}}=e^{-\alpha\sum_{\Bracket{i,j}}(s_{i}-s_{j})^{2}}, & \alpha>0
\end{array}
\end{equation}
Then the correlation function becomes 
\begin{equation}
C_{ij}^{(A)}\sim e^{-|i-j|/\xi},\;\;\; \xi>0
\end{equation}
where $\xi$ is the correlation length, which is considered finite and short, ensuring locality. 
The emergent metric from correlations $C_{ij}^{A}$ is approximately Minkowski-like
\begin{equation}
ds^{2}=-c^{2}dt^{2}+dx^{2}
\end{equation}
For the case of the semiclassical limit B (with causal loops) we choose a semiclassical limit emphasising non-local cyclic correlations. Consider now a quantum state whose correlations are cyclic and nonlocal. 
Define the state as a cyclic coherent superposition
\begin{equation}
\ket{\Psi}_{loop}=\frac{1}{\sqrt{N}}\sum_{k=1}^{N}e^{i2\pi k/N}\ket{s_{k}},\;\; \ket{s_{N+1}}=\ket{s_{1}}
\end{equation}
This cyclic superposition explicitly has correlations wrapping around 
\begin{equation}
C_{ij}^{B}=\bra{\Psi_{loop}}\hat{S}_{i}^{z}\hat{S}_{j}^{z}\ket{\Psi_{loop}}\sim cos(\frac{2\pi(i-j)}{N})
\end{equation}
(note the cyclic periodicity $C_{i,i+N}^{(B)}=C_{i,i}^{(B)}$). 
The emergent geometry now includes closed loops 
\begin{equation}
ds^{2}\sim -c^{2}dt^{2}+R^{2}d\theta^{2},\;\; 0\leq\theta<2\pi
\end{equation}
where $\theta$ is periodic giving explicitly a closed spatial dimension. Closed loops allow events to causally influence their own past. 
As an example, consider $N=4$ spins. In the Minkowski limit, the state coefficients are short range and we have 
\begin{equation}
\ket{\Psi_{A}}\sim\ket{\uparrow\uparrow\downarrow\downarrow}+\ket{\downarrow\downarrow\uparrow\uparrow}
\end{equation}
the correlation function is 
\begin{equation}
C^{(A)}_{ij}\sim e^{|i-j|/\xi},\;\;\xi\sim 1
\end{equation}
which clearly decays with distance. In the limit $B$, the cyclic quantum state is 
\begin{equation}
\ket{\Psi_{B}}=\frac{1}{2}(\ket{\uparrow\downarrow\downarrow\uparrow}+\ket{\downarrow\uparrow\uparrow\downarrow}+\ket{\uparrow\uparrow\downarrow\downarrow}+\ket{\downarrow\downarrow\uparrow\uparrow})
\end{equation}
and the correlation function is cyclic
\begin{equation}
C_{ij}^{(B)}\sim cos(\frac{\pi(i-j)}{2})
\end{equation}
We clearly see that in the first case the correlations are short range, stable and robust. Small perturbations preserve Minkowski structure. In the second case, correlations are cyclic, nonlocal and a delicate coherence is needed. Such loops may easily decohere or become unstable with perturbations. But under specific conditions, such as protected topological phases, these cyclic correlations could become robust and indeed this is plausible in quantum spin liquids with topological order. 
To see exactly how the metric tensor emerges from correlations, let's consider the discrete correlation $C_{ij}$ and write it as a continuous function
\begin{equation}
C(x,y)=\Bracket{\hat{\mathcal{O}}(x)\hat{\mathcal{O}}(y)}-\Bracket{\hat{\mathcal{O}}(x)}\Bracket{\hat{\mathcal{O}}(x)}
\end{equation}
In the continuum limit, define the metric tensor $g_{\mu\nu}(x)$ from infinitesimal distances
\begin{equation}
d(x,x+dx)^{2}=g_{\mu\nu}(x)dx^{\mu}dx^{\nu}
\end{equation}
For very close points we have infinitesimal correlations 
\begin{equation}
C(x,x+dx)\sim C_{0}e^{-g_{\mu\nu}(x)dx^{\mu}dx^{\nu}}
\end{equation}
or, by expanding for small separations
\begin{equation}
C(x,x+dx)\sim C_{0}(1-g_{\mu\nu}(x)dx^{\mu}dx^{\nu})
\end{equation}
and therefore the emergent metric tensor is 
\begin{equation}
g_{\mu\nu}(x)=-\frac{1}{C_{0}}\frac{\partial^{2}C(x,y)}{\partial x^{\mu}\partial y^{\nu}}_{(y\rightarrow x)}
\end{equation}
For a closed causal loop 
\begin{equation}
C(\theta)=C_{0}cos(\frac{\theta}{R}),\;\;0\leq\theta <2\pi R
\end{equation}
Expanding around small distances
\begin{equation}
C(\theta)\sim C_{0}(1-\frac{\theta^{2}}{2 R^{2}})
\end{equation}
and the emergent metric for small $\theta$ is 
\begin{equation}
ds^{2}\sim \frac{d\theta^{2}}{R^{2}}
\end{equation}
where the geometry clearly has closed loops. This is a simple example. 
In practice, achieving quantum states that can give rise to emergent exotic causal structures (for example closed causal loops) requires quantum systems known for their unusual and stable long range correlations such as quantum spin liquids with high entangled phases of matter exhibiting topological order and robust non-local correlations, or fractional quantum hall states with topological protection and robust cyclic correlations. In such experiments, given the assumption that quantum correlations define geometry itself, if sufficiently robust correlations exist, then the stable large scale quantum correlations could produce an effective emergent geometry. In this case quantum matter with cyclic or non-local correlations might create an effective geometry with exotic causal structure (like closed loops) in its internal emergent spacetime. To an observer whose measurement devices are defined by interactions within this quantum state, the internal geometry can exhibit causal loops due to quantum correlations. They measure effective geometry shaped by quantum coherence. An external observer in the surrounding Minkowski space would still see the geometry around the quantum matter as standard. To the external observer, the exotic causal structure inside the quantum matter might appear as unusual transport or correlation properties. In spin liquid systems this would amount to an internal geometry where observers embedded in that quantum state would effectively experience exotic geometry. This is very different from the situation in which exotic geometry were to be created via modifications of correlations among quantum modes that make up light, as a classical light would represent an open system that would redefine the spacetime geometry locally for any observer using any type of measuring devices. The disturbance would still be local in the sense of affecting a limited spacetime region, but would be detectable by observers using any type of measuring device, not necessarily measuring devices embedded in the modified quantum system. 
\section{Uhlmann gauge theory stabilisation of closed causal loops}
Let us now start with constructing a quantum system that explicitly has quantum correlations corresponding to a closed causal loop. We allow this system to decohere so that we obtain a semiclassical mixed state (a density matrix) corresponding to a classical geometry possibly containing a closed causal loop (but typically fragile due to further decoherence). We purify this semiclassical mixed state explicitly by introducing ancillary degrees of freedom. After purification, the total system (original + ancillas) is in a pure state, encoding the classical coherence structure as an entanglement between ancillas and semiclassical modes. 
By adjusting the coherence structure explicitly through Uhlmann gauge transformations, we can then stabilise the coherence and entanglement structure obtaining a new robust semiclassical state explicitly corresponding to a stable closed timelike loop geometry. Let us therefore start with a mixed quantum state that contains closed-loop correlations
\begin{equation}
\rho_{initial}=\sum_{i,j}\rho_{ij}\ket{i}\bra{j}
\end{equation}
we allow for decoherence and arrive at a semiclassical density matrix
\begin{equation}
\rho_{sc}=\sum_{n}p_{n}\ket{n}\bra{n},\;\; p_{n}>0,\;\; \sum_{n}p_{n}=1
\end{equation}
Here $\{\ket{n}\}$ is a preferred classical like basis corresponding explicitly to classical geometry modes with loops. 
We then purify $\rho_{sc}$ by introducing ancilla Hilbert space $\mathcal{H}_{A}$
\begin{equation}
\ket{\Psi}=\sum_{n}\sqrt{p_{n}}\ket{n}_{system}\otimes \ket{a_{n}}_{ancilla}
\end{equation}
The purified state is in enlarged Hilbert space and will produce after tracing over the ancillas the original semiclassical state 
\begin{equation}
\rho_{sc}=Tr_{A}(\ket{\Psi}\bra{\Psi})
\end{equation}
The purification encodes the classical coherence in the form of entanglement between ancilla and system. 
Next, we use an Uhlmann gauge approach in this space of purifications, to modify the coherence structure. We define, following the ideas behind the construction of a dynamical Uhlmann gauge theory, the Uhlmann gauge fields $A_{\mu}$ defined through the Uhlmann parallel transport of purified states 
\begin{equation}
A_{\mu}(\rho)=\rho^{-1/2}\partial_{\mu}\rho^{1/2}
\end{equation}
The Uhlmann charges define quantum misalignments or coherent phase charges. They can be represented by 
\begin{equation}
Q_{Uhlmann}=\bra{\Psi}\mathcal{G}\ket{\Psi}
\end{equation}
where $\ket{\Psi}$ is the purification and $\mathcal{G}$ is the gauge generator. 
One can explicitly re-write this as 
\begin{equation}
Q_{Uhlmann}=\rho^{1/2}[A_{\mu},A_{\nu}]\rho^{1/2}
\end{equation}
The Uhlmann gauge transformations modify the phases and coherence structure in the purified state
\begin{equation}
\begin{array}{cc}
\ket{\Psi}\rightarrow\ket{\Psi'}=U(\rho)\ket{\Psi},& U(\rho)=\mathcal{P} exp(i\oint A_{\mu}dx^{\mu})
\end{array}
\end{equation}
(note that here the indices refer to the Uhlmann gauge bundle and hence to transformations in the space of purification, where confusion can occur, I will change the notation)
We can apply the gauge transformations to stabilise closed-loop coherence. We choose the Uhlmann gauge transformation $U(\rho)$ that imposes non-trivial holonomy
\begin{equation}
U(\rho)=\mathcal{P}exp(i\oint_{C_{loop}}A_{\mu}dx^{\mu})\neq 1
\end{equation}
such a non-trivial holonomy stabilises cyclic coherence loops geometrically. Small fluctuations away from the stable loop induce gauge fields restoring the original stable coherence. Thus, cyclic coherence becomes robust and protected against decoherence through our gauge structure. 
After applying the Uhlmann gauge stabilisation, the new purified state is 
\begin{equation}
\ket{\Psi_{stable}}=U(\rho)\ket{\Psi}=\sum_{n}\sqrt{p_{n}}e^{i\phi_{n}}\ket{n}_{system}\otimes \ket{a_{n}}_{ancilla}
\end{equation}
Gauge phases $e^{i\phi_n}$ have been chosen to stabilise coherence loops. 
Taking the partial trace explicitly then produces a stabilised semiclassical density matrix
\begin{equation}
\rho_{stable}=Tr_{A}(\ket{\Psi_{stable}}\bra{\Psi_{stable}})=\sum_{n,m}\sqrt{p_{n}p_{m}}e^{i(\phi_{n}-\phi_{m})}\Bracket{a_{m}|a_{n}}\ket{n}\bra{m}
\end{equation}
This would generate the correlations required for the stabilised closed loop geometry. 
Therefore the quantum coherence and entanglement is protected via the Uhlmann method where non-trivial Uhlmann holonomies explicitly introduced by means of shifting the phase structure of the ancillas resulted in robustness of the cyclic correlations. 

\section{conclusions}
Several aspects were discussed in this article. First, I introduced coherent string states from a more realistic point of view, focused on phase consistency conditions and not on states being eigenvalues of the annihilation operators. In this context it was possible to construct coherent states in an approximate sense and to preserve some quantum properties of D-branes. I reconstructed worldsheet-target space duality in this context and I derived notions like string length from a strictly quantum coherence perspective, abandoning any geometrical assumptions. This method allowed the introduction of mathematical tools that allowed us to maintain quantum properties of D-branes. This may open the way towards a modified AdS/CFT duality that will contain semiclassical D-branes at the level of the pre-Maldacena limit of the gauge sector. In this way, the resulting duality may include automatically quantum properties that may be translated in the AdS bulk that was previously considered to be strictly speaking classical. 
Next, I derived a new method of stabilising closed temporal loops and showed some simple toy-models associated to it. The method is based on the Uhlmann gauge approach which I presented briefly.


\begin{thebibliography}{99}
\bibitem{1}J. Maldacena, Int. J. Theor. Phys. Vol 38, Pag. 1113 (1999)
\bibitem{2}J. Khoury, H. Verlinde, Vol. 3,  No. 6, Pag. 1893 (1999)
\bibitem{3}H. Shin, K. Sugiyama, K. Yoshida, Nucl. Phys. B Vol. 6, Iss. 1-2, Pag. 78 (2003)
\bibitem{4}S. Ebert, H. Y. Sun, Z. Yan, JHEP Vol. 2022, No. 161 (2022)
\bibitem{5}P. Vanhove, F. Zerbini, PoS 383, 10.22323/1383.0022 (2021)
\bibitem{6}K. Hashimoto, W. Taylor, JHEP 0310, No. 040 (2003)
\bibitem{7}O. DeWolfe, P. Romatschke, JHEP 2019, No. 272 (2019)
\bibitem{8}J. Hansen, B. H. Lee, C. Park, D. H. Yeom, Class. Quantum Grav. 30, 235022 (2013)
\bibitem{9}N. Kajuri, SciPost Phys. Lect. Notes 22 (2021)
\bibitem{10}X. Dong, D. Harlow, A. C. Wall, Phys. Rev. Lett. 117, 021601 (2016)
\bibitem{11}K. Papadodimas, S. Raju, Phys. Rev. D 89, 086010 (2014)
\bibitem{12}J. L. Karczmarek, C. Rabideau, JHEP Vol. 2013, No. 78 (2013)
\end{thebibliography}
\end{document}